\documentclass{wsmodified}

\newcommand{\be}{\begin{equation}}
\newcommand{\bea}{\begin{eqnarray}}
\newcommand{\en}{\end{equation}}
\newcommand{\eea}{\end{eqnarray}}

\newcommand{\fract}[2]{{\textstyle\frac{#1}{#2}}} 

\begin{document}

\title{Pentaquarks and Radially Excited 
Baryons\footnote{\uppercase{C}ontribution to the workshop 
\uppercase{NSTAR} 2004,\uppercase{G}renoble, \uppercase{M}arch 2004.}}

\author{H. Weigel}

\address{Fachbereich Physik, Siegen University\\
Walter--Flex--Stra{\ss}e 3, D--57068 Siegen, Germany}

\maketitle

\abstracts{In this talk I report on a computation of
the spectra of exotic pentaquarks and radial excitations
of the low--lying $\frac{1}{2}^+$ and $\frac{3}{2}^+$ baryons in
a chiral soliton model.}

\section{Introduction}

Although chiral soliton model predictions for the mass of the 
lightest exotic pentaquark, the $\Theta^+$ with zero isospin and unit 
strangeness, have been around for some time\cite{early}, the study 
of pentaquarks as baryon resonances became popular only 
recently when experiments\cite{Thetaexp,NA49} indicated their 
existence. These experiments were stimulated by a chiral soliton model 
estimate\cite{Di97} suggesting that such exotic baryons might 
have a width\footnote{In chiral soliton models the direct extraction of
the interaction Hamiltonian for hadronic decays of resonances still 
is an unresolved issue. Estimates are obtained from axial current 
matrix elements\cite{Di97,We98,Pr03,Ja04}. In view of what is 
known about the related $\Delta\to\pi N$ transition matrix 
element\cite{deltadecay}, such estimates may be questioned.}
so small\cite{Di97,Ja04} that it could have escaped earlier detection.
Then very quickly the novel observations initiated exhaustive 
studies on the properties of pentaquarks. Comprehensive 
lists of such studies are, for example, collected in 
refs.\cite{Je03,El04}.

\begin{figure}[t]
\centerline{
\parbox[t]{4cm}{\epsfxsize=3.2cm\epsfbox{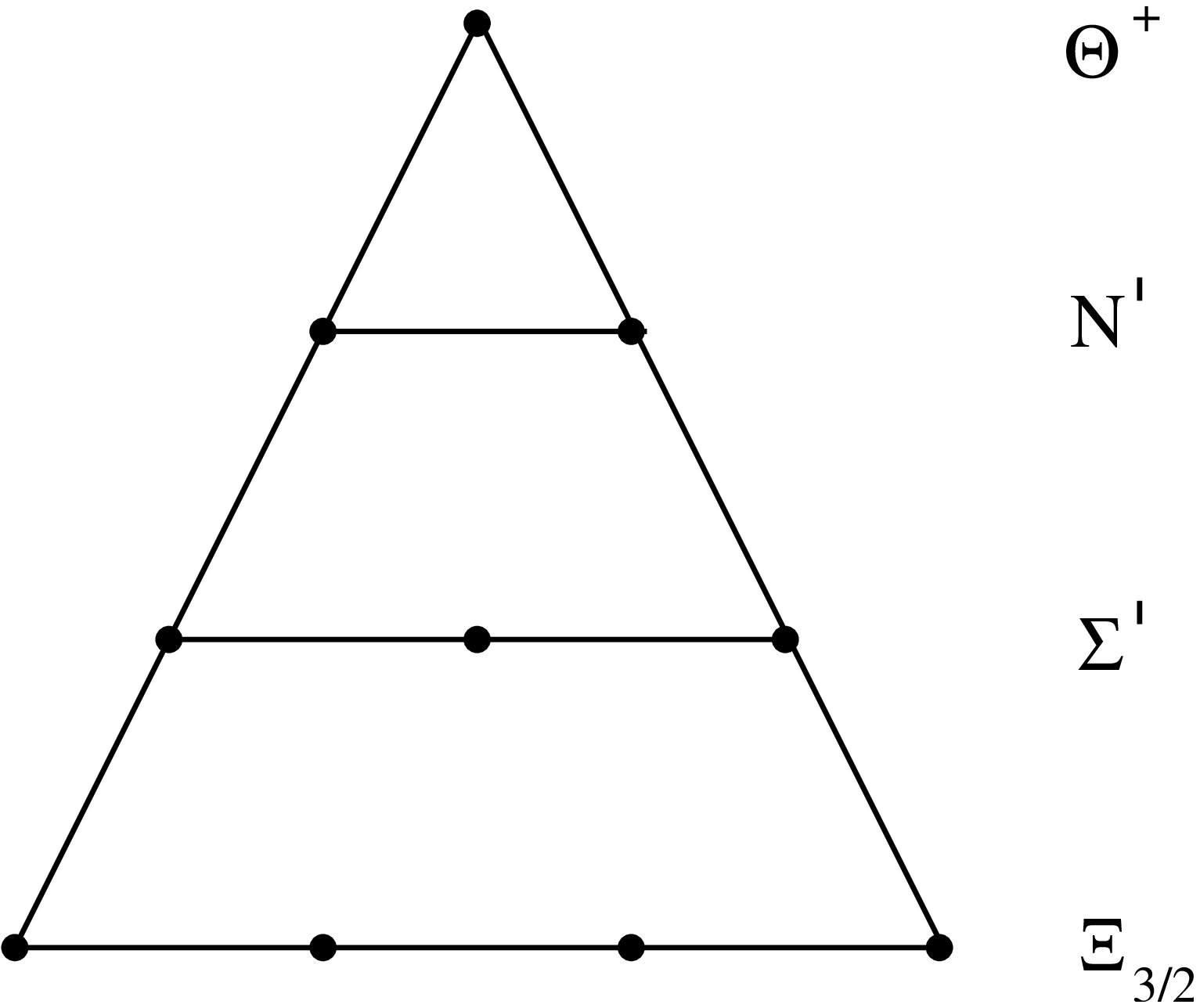}}~~~
\parbox[t]{4cm}{\vskip-3.4cm\epsfxsize=4.5cm\epsfbox{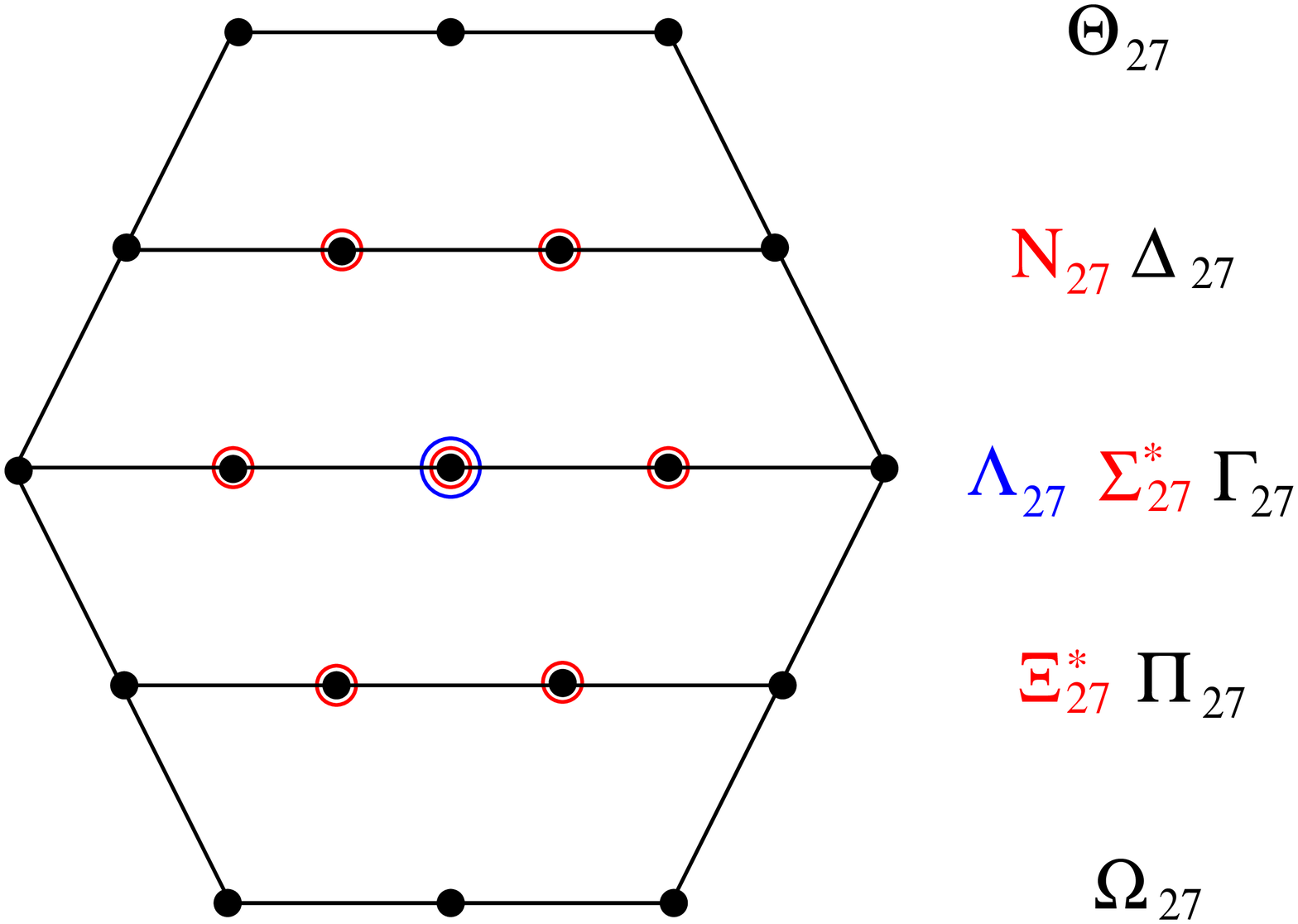}}}
\caption{\label{fig_1}A sketch of the $SU_F(3)$ representations
$\overline{\mathbf{10}}$ and $\mathbf{27}$ with their 
exotic baryon members: $\Theta^+,\Xi_{3/2},\Theta_{27},
\Lambda_{27},\Gamma_{27},\Pi_{27}$ and~$\Omega_{27}$.
As usual various isospin projections are plotted horizontally 
while states of different hypercharge are spread vertically.}
\vskip-1.0cm
\end{figure}

In chiral soliton models states with baryon quantum numbers 
are generated from the soliton by canonically quantizing the 
collective coordinates associated with (would--be) zero modes 
such as $SU(3)$ flavor rotations. The lowest states are 
members of the flavor octet ($J^\pi=\fract{1}{2}^+$) and 
decuplet representations ($J^\pi=\fract{3}{2}^+$).
Due to flavor symmetry breaking the 
physical states acquire admixtures from higher dimensional 
representations. For the $J^\pi=\frac{1}{2}^+$ baryons those 
admixtures originate dominantly from the antidecuplet, 
$\overline{\mathbf{10}}$, and the 
$\mathbf{27}$--plet\cite{Pa89}. The particle 
content of these representations is depicted in figure~\ref{fig_1}.
They also contain states with quantum numbers 
that \emph{cannot} be built as three--quark composites but
contain additional quark--antiquark pairs. 
Hence the notion of exotic pentaquarks.
So far, the $\Theta^+$ and $\Xi_{3/2}$ with masses of 
$1537\pm10{\rm MeV}$~\cite{Thetaexp} and $1862\pm2{\rm MeV}$~\cite{NA49}
have been observed, however, the single observation of 
$\Xi_{3/2}$ is not undisputed\cite{Fi04}. Soliton models 
predict the quantum numbers $I(J^\pi)=0(\frac{1}{2}^+)$ for 
$\Theta^+$ and $\frac{3}{2}(\frac{1}{2}^+)$ for~$\Xi_{3/2}$. 
Experimentally these quantum numbers are not yet confirmed.

It is natural to wonder about the nature of states in 
higher dimensional representations that do carry quantum numbers
of three--quark composites, such as the $N^\prime$
and $\Sigma^\prime$ in figure~\ref{fig_1}. Radial 
excitations\cite{Ha84} of the octet nucleon and $\Sigma$ are expected 
to have masses similar to $N^\prime$ and $\Sigma^\prime$. Hence 
sizable mixing should occur between the antidecuplet and an octet 
of radial excitations. In a rough approximation this corresponds
to the picture that pentaquarks are members of the direct sum 
$\mathbf{8}\oplus\overline{\mathbf{10}}$ which is also obtained in 
a quark--diquark approach\cite{Ja03}. The possible
mixing between radially excited octet and antidecuplet baryons was
recognized already earlier\cite{Sch91,Sch91a} showing that
the full picture is more complicated. A dynamical model was 
developed\cite{Sch91,Sch91a} to investigate such mixing effects and 
also to describe static properties of the low--lying 
$J^\pi=\frac{1}{2}^+$ and 
$J^\pi=\frac{3}{2}^+$ baryons. Essentially that model has only a 
single free parameter, the Skyrme constant $e$ which should be in 
the range $e\approx 5.0\ldots5.5$. Later the mass of 
the recently discovered $\Theta^+$ pentaquark was predicted with 
reasonable accuracy in the same model\cite{We98}. In this talk
I will present predictions for masses of the $\Xi_{3/2}$ and additional 
exotic baryons that originate the $\mathbf{27}$--plet from 
exactly that model without any further modifications. The
latter may be considered as partners of $\Theta^+$ and $\Xi_{3/2}$ in 
the same way as the $\Delta$ is the partner of the nucleon. 

In section 2 of this talk I will review the 
quantization of the rotational and radial degrees of freedom
of a chiral soliton. I will present numerical results on the 
spectrum in section 3 and summarize in section 4. 
A complete description of the material presented in this
talk may be found in ref.\cite{We04}.

\section{Collective Quantization of the Soliton}

I consider a chiral Lagrangian in flavor $SU(3)$. The basic
variable is the chiral field $U={\rm exp}(i\lambda_a\phi^a/2)$
that represents the pseudoscalar fields $\phi^a$ ($a=0,\ldots,8$). 
Other fields may be included as well. For example, the specific model 
used later also contains a scalar meson. In general a chiral 
Lagrangian can be decomposed as a sum,
${\mathcal L}={\mathcal L}_{\rm S}+{\mathcal L}_{\rm SB}$,
of flavor symmetric and flavor symmetry breaking pieces.
Denoting the (classical) soliton solution of this Lagrangian by 
$U_0(\vec{r})$ states with baryon quantum numbers are 
constructed by quantizing the flavor rotations 
\be
U(\vec{r},t)=A(t)U_0(\vec{r})A^\dagger(t), \qquad
A(t)\in SU(3)
\label{collcor1}
\en
canonically. According to the above separation the Hamiltonian for the 
collective coordinates $A(t)$ can be written as $H=H_{\rm S}+H_{\rm SB}$.
For unit baryon number the eigenstates of $H_{\rm S}$ are the members 
of $SU(3)$ representations with the condition that the representation
contains a state with identical spin and isospin quantum numbers
(such as {\it e.g.} the nucleon or the $\Delta$). 
In order to include radial excitations that potentially
mix with states in higher dimensional $SU(3)$ representations
the corresponding collective coordinate~$\xi(t)$ is introduced
via\cite{Ha84,Sch91,Sch91a}
\be
U(\vec{r},t)=A(t)U_0(\xi(t)\vec{r})A^\dagger(t)\, .
\label{collcor2}
\en
Changing to $x(t)=[\xi(t)]^{-3/2}$ the flavor symmetric
piece of the collective Hamiltonian for a given $SU(3)$ 
representation of dimension $\mu$ reads
\be
H_{\rm S}=\frac{-1}{2\sqrt{m\alpha^3\beta^4}}\frac{\partial}{\partial x}
\sqrt{\frac{\alpha^3\beta^4}{m}}\frac{\partial}{\partial x}
+V+\left(\frac{1}{2\alpha}-\frac{1}{2\beta}\right)J(J+1)
+\frac{1}{2\beta}C_2(\mu)+s\, ,
\label{freebreath}
\en
where $J$ and $C_2(\mu)$ are the spin and (quadratic) Casimir
eigenvalues associated with the representation $\mu$. 
Note that $m=m(x),\alpha=\alpha(x),\ldots,s=s(x)$ are functions of
the scaling variable to be computed in the specified soliton
model\cite{Sch91,Sch91a}. For a prescribed $\mu$ there are 
discrete eigenvalues~(${\mathcal E}_{\mu,n_\mu}$) and 
eigenstates~($|\mu,n_\mu\rangle$) of $H_{\rm S}$.
The radial quantum number $n_\mu$ counts the number
of nodes in the respective wave--functions.
I now employ these to compute matrix elements of
the full Hamiltonian
\be
H_{\mu,n_\mu;\mu^\prime,n^\prime_{\mu^\prime}}=
{\mathcal E}_{\mu,n_\mu}\delta_{\mu,\mu^\prime}
\delta_{n_\mu,n^\prime_{\mu^\prime}}
-\langle\mu,n_\mu 
|\fract{1}{2}{\rm tr}\left(\lambda_8A\lambda_8A^\dagger\right)
s(x)|\mu^\prime,n^\prime_{\mu^\prime}\rangle \, .
\label{hammatr}
\en
This ``matrix'' is diagonlized exactly (rather than in some
approximation scheme) yielding the baryonic states
$|B,m\rangle=\sum_{\mu,n_\mu}C_{\mu,n_\mu}^{(B,m)}
|\mu,n_\mu\rangle \,$. Here $B$ refers to the specific 
baryon and $m$ labels its excitations.
Before presenting numerical results for the 
spectrum of the Hamiltonian~(\ref{hammatr}) I would like to
stress that quantizing the radial degree of 
freedom is also demanded by observing that the proper description 
of baryon magnetic moments
requires a substantial feedback of flavor symmetry breaking on
the soliton size\cite{Schw91}.

\section{Results}

I divide the model results into three categories. First there 
are the low--lying $J=\frac{1}{2}$ and $J=\frac{3}{2}$ 
baryons together with their monopole excitations. Without flavor 
symmetry breaking these would be pure octet and decuplet states. 
Second are the $J=\frac{1}{2}$ states that are
dominantly members of the antidecuplet. Those that are non--exotic
mix with octet baryons and their monopole excitations. 
Third are the $J=\frac{3}{2}$ baryons that would dwell in the 
$\mathbf{27}$--plet if flavor symmetry held.
The $J=\frac{1}{2}$ baryons from the $\mathbf{27}$--plet are heavier 
than those with $J=\frac{3}{2}$ and will thus not be studied here. 

\subsection{Ordinary Baryons and their Monopole Excitations}

Table \ref{tab_1} shows the predictions for the mass differences 
with respect to the nucleon of the eigenstates of the full 
Hamiltonian~(\ref{hammatr}) for two values of the Skyrme 
parameter $e$. 
\begin{table}[t]
\tbl{\label{tab_1}
Mass differences of the eigenstates of the Hamiltonian 
(\protect\ref{hammatr}) with respect to the nucleon in MeV. Experimental 
data\protect\cite{PDG02} refer to four and three star resonances, 
unless otherwise noted. For the Roper resonance [$N(1440)$] 
I list the Breit--Wigner~(BW) mass and the pole position~(PP) 
estimate\protect\cite{PDG02}. The states~"?" are potential
isospin $\frac{1}{2}$~$\Xi$ candidates with yet undetermined
spin--party.} 
{\hspace{-0.3cm}\begin{tabular}{c | c c c | c c c | c c c}
B& \multicolumn{3}{c|}{$m=0$} & \multicolumn{3}{c|}{$m=1$}
& \multicolumn{3}{c}{$m=2$} \\
\hline
& $e$=5.0 & \hspace{-0.2cm}$e$=5.5 & \hspace{-0.2cm}expt. 
& $e$=5.0 & \hspace{-0.2cm}$e$=5.5 & \hspace{-0.2cm}expt.
& $e$=5.0 & \hspace{-0.2cm}$e$=5.5 & \hspace{-0.2cm}expt. \\ 
\hline
\vspace{-0.1cm}
&&&&&&&&\\
N & \multicolumn{3}{c|}{Input}
            & 413 & 445 & 
\parbox[t]{1.1cm}{\vskip-0.4cm\large${\rm 501~BW}\atop{\rm 426~PP}$}
  & 836 & 869 & 771  \\
$\Lambda$ 
& 175 & 173 & 177 & 657 & 688 & 661 & 1081 & 1129 & 871 \\
\vspace{-0.3cm}
&&&&&&&&\\
$\Sigma$
& 284 & 284 & 254 & 694 & 722 & 721 & 1068 & 1096 & 
\parbox[t]{0.6cm}{\vskip-0.35cm
\large${\rm 831~(*)~}\atop{\rm 941~(**)}$}
\\
\vspace{-0.3cm}
&&&&&&&&\\
$\Xi$
& 382 & 380 & 379 & 941 & 971 & 
\parbox[t]{0.7cm}{\vskip-0.35cm\large${\rm 751}\atop{\rm 1011}$}(?)
& 1515 & 1324 & --- \\
\vspace{-0.3cm}
&&&&&&&&\\
\hline
$\Delta$
& 258 & 276 & 293 & 640 & 680 & 661 & 974 & 1010 & 981 \\
$\Sigma^*$
& 445 & 460 & 446 & 841 & 878 & 901 & 1112 & 1148 & 1141 \\
$\Xi^*$
& 604 & 617 & 591 & 1036 & 1068 & --- & 1232 & 1269 & --- \\
$\Omega$
& 730 & 745 & 733 & 1343 & 1386 & --- & 1663 & 1719 & --- \\
\end{tabular}}
\end{table}
The agreement with the experimental data is quite astonishing. Only
the Roper resonance ($|N,1\rangle$) is predicted a bit on the 
low side when compared to the empirical Breit--Wigner mass but 
agrees with the estimated pole position\footnote{For 
other resonances the discrepancy between the Breit--Wigner mass and the 
pole position is much smaller and there is no need to distinguish 
between them.}. This is common for the breathing mode approach 
in soliton models\cite{Ha84}.  All other first
excited states are quite well reproduced. For the $\frac{1}{2}^+$ 
baryons the energy eigenvalues for the second excitations overestimate 
the corresponding empirical data somewhat. In the nucleon channel 
the model predicts the $m=3$ state 
only about $40{\rm MeV}$ higher than the $m=2$ state, {\it i.e.} 
still within the regime where the model is assumed to be applicable. 
This is interesting because empirically it is suggestive that there 
might exist more than only one resonance in that energy
region\cite{Ba95}. For the $\frac{3}{2}^+$ baryons with $m=2$ the 
agreement with data is on the 3\% level. The particle 
data group\cite{PDG02} lists two ``three star'' isospin--$\frac{1}{2}$ 
$\Xi$ resonances at $751$ and $1011{\rm MeV}$ above the nucleon
whose spin--parity is not yet determined. The present model suggests 
that the latter is $J^\pi=\frac{1}{2}^+$, while the former seems to 
belong to a different channel. 

The present model gives fair agreement with available data and thus
supports the picture of coupled monopole and rotational modes. 
Most importantly, the inclusion of higher 
dimensional $SU_F(3)$ flavor representations in three flavor 
chiral models does \emph{not} lead to the prediction of any novel 
states in the regime between $1$ and $2{\rm GeV}$ in the 
non--exotic channels.

\subsection{Exotic Baryons from the Antidecuplet}

Table~\ref{tab_2} compares the model prediction for the exotics  
$\Theta^+$ and $\Xi_{3/2}$ to available data\cite{Thetaexp,NA49} and to 
a chiral soliton model calculation\cite{Wa03} that does not include 
a dynamical treatment of the monopole excitation. In that calculation 
parameters have been tuned to reproduce the mass of the lightest
exotic pentaquark, $\Theta^+$. The inclusion of the monopole
excitation increases the mass of the $\Xi_{3/2}$ slightly
and brings it closer to the empirical value. Furthermore,
the first prediction\cite{Di97} for the mass of the $\Xi_{3/2}$ 
was based on identifying $N(1710)$ with the nucleon like state in the 
antidecuplet and thus resulted in a far too large mass of $2070{\rm MeV}$. 
Other chiral soliton model studies 
either take $M_{\Xi_{3/2}}$ as input\cite{Bo03}, 
adopt the assumptions of ref.~\cite{Di97} or are less predictive because 
the model parameters vary considerably\cite{El04}. 

\begin{table}[t]
\tbl{\label{tab_2}
Masses of the eigenstates of the Hamiltonian (\ref{hammatr}) 
for the exotic baryons $\Theta^+$ and~$\Xi_{3/2}$. The absolute
energy scale is set by the nucleon mass.
Experimental data are the average of refs.\protect\cite{Thetaexp} 
for $\Theta^+$ and the NA49 result for $\Xi_{3/2}$\protect\cite{NA49}.
I also compare the predictions for the ground state ($m=0$) to the 
treatment of ref.\protect\cite{Wa03}. All energies are in~GeV.} 
{\begin{tabular}{c | c c c |c | c c c}
B& \multicolumn{4}{c|}{$m=0$} & \multicolumn{3}{c}{$m=1$}\\
\hline
& $e$=5.0 & $e$=5.5 & expt. & WK\cite{Wa03}
& $e$=5.0 & $e$=5.5 & expt. \\
\hline
$\Theta^+$ & $1.57$ & $1.59$ & $1.537\pm0.010$ & $1.54$ &
$2.02$ & $2.07$ & -- \\
$\Xi_{3/2}$ & $1.89$ & $1.91$ & $1.862\pm0.002$ & $1.78$ &
$2.29$ & $2.33$ & -- 
\end{tabular}}
\end{table}
Without any fine--tuning the model prediction is only about 
$30$--$50{\rm MeV}$ higher than the data. In view of the approximative 
nature of the model this should be viewed as good agreement. 
Especially the mass difference between the two potentially observed 
exotics is reproduced within $10{\rm MeV}$.

\subsection{Baryons from the 27--plet}

As seen from figure~\ref{fig_1}, the $\mathbf{27}$--plet
contains states with the quantum numbers of the baryons
that are also contained in the decuplet of the low--lying
$J=\frac{3}{2}$ baryons: $\Delta,\Sigma^*$ and $\Xi^*$. 
Under flavor symmetry breaking these states mix with the radial 
excitations of decuplet baryons and are already discussed in
table~\ref{tab_1}. Table~\ref{tab_3} shows the model predictions 
for the $J=\frac{3}{2}$ baryons that emerge from the $\mathbf{27}$--plet
but do not have partners in the decuplet. Again, the experimental 
nucleon mass is used to set the mass scale.
\begin{table}[t]
\tbl{\label{tab_3}
Predicted masses 
of the eigenstates of the Hamiltonian (\protect\ref{hammatr}) for
the exotic $J=\frac{3}{2}$ baryons with $m=0$ and $m=1$ that originate
from the $\mathbf{27}$--plet with hypercharge (Y) and isospin (I) 
quantum numbers listed. I also compare the $m=0$ case to treatments of 
refs.\protect\cite{Wa03,Bo03,Wu03}. All numbers are in GeV.}
{\begin{tabular}{c | c c | c c |c|c|c|cc}
\multicolumn{3}{c|}{B}& \multicolumn{5}{c|}{$m=0$} 
& \multicolumn{2}{c}{$m=1$}\\
\hline
& Y~ & ~I & ~$e$=5.0 & $e$=5.5~ 
& WK\cite{Wa03} & BFK\cite{Bo03} & WM\cite{Wu03} 
&~$e$=5.0 & $e$=5.5~ \\
\hline
$\Theta_{27}$ &$2$ & $1$ & $1.66$ & $1.69$ & $1.67$ 
& $1.60$ & $1.60$ & 2.10 & 2.14\\
$N_{27}$ &$1$ & $1/2$ & $1.82$ & $1.84$ & $1.76$ 
& $ -- $ & $1.73$ & 2.28 & 2.33  \\
$\Lambda_{27}$ &$0$ & $0$ & $1.95$ & $1.98$ & $1.86$ 
& $ -- $ & $1.86$ & 2.50 & 2.56\\
$\Gamma_{27}$ &$0$ & $2$ & $1.70$ & $1.73$ & $1.70$ 
&  1.70  & $1.68$ & 2.12 & 2.17 \\
$\Pi_{27}$ &$-1$ & $3/2$ & $1.90$ & $1.92$ & $1.84$ 
& $1.88$ & $1.87$ & 2.35 & 2.40\\
$\Omega_{27}$ &$-2$ & $1$ & $2.08$ & $2.10$ & $1.99$ 
& $2.06$ & $2.07$ & 2.54 & 2.59
\end{tabular}}
\end{table}
Let me remark that the particle data group\cite{PDG02} lists two states 
with the quantum numbers of $N_{27}$ and $\Lambda_{27}$ at $1.72$ 
and $1.89{\rm GeV}$, respectively, that fit reasonably well into the 
model calculation. In all channels the $m=1$ states turn out to be 
about $500{\rm MeV}$ heavier than the exotic ground states.
When combined with the $m=1$ states of $\Delta$, $\Sigma^*$, 
and $\Xi^*$ in table~\ref{tab_1} it is observed
that the masses of states that are degenerate in
hypercharge decrease with isospin, that is
$M_{|\Delta,1\rangle}<M_{|N_{27},0\rangle}$,
$M_{|\Gamma_{27},0\rangle}<M_{|\Sigma^*,1\rangle}
<M_{|\Lambda_{27},0\rangle}$,
$M_{|\Pi_{27},0\rangle}<M_{|\Xi^*,1\rangle}$, and
$M_{|\Omega_{27},0\rangle}<M_{|\Omega,1\rangle}$.

\section{Conclusion}

In this talk I have discussed the interplay between rotational and 
monopole excitations for the spectrum of pentaquarks in a chiral 
soliton model. In this approach the scaling degree of freedom has 
been elevated to a dynamical quantity which has been quantized 
canonically at the same footing as the (flavor) rotational modes.
Then not only the ground states in individual irreducible 
$SU_F(3)$ representations are eigenstates of the (flavor--symmetric 
part of the) Hamiltonian but also all their radial excitations. 
I have treated flavor symmetry breaking 
exactly rather then only at first order. Thus, even though the chiral 
soliton approach initiates from a flavor--symmetric formulation, 
it is capable of accounting for large deviations thereof. 

The spectrum of the low--lying $\frac{1}{2}^+$ and $\frac{3}{2}^+$ 
baryons is reasonably well reproduced. Also, the model results for 
various static properties are in acceptable agreement with the 
empirical data\cite{Sch91}. This makes the model reliable to 
study the spectrum of the excited states. Indeed the model 
states can clearly be identified with observed baryon excitations; 
except maybe an additional P11 nucleon state although there exist
analyses with such a resonance. Otherwise, this model calculation
did not indicate the existence of yet unobserved baryon
states with quantum numbers of three--quark composites.
The computed masses for the exotic $\Theta^+$
and $\Xi_{3/2}$ baryons nicely agree with the recent observation
for these pentaquarks. At this stage the model contains no more 
adjustable parameter. The mass difference between mainly octet
and mainly antidecuplet baryons thus is a prediction while it is an 
input quantity in most other approaches\cite{Di97,Wa03,Bo03,Wu03,El04}.
Thus the present predictions for the masses of the spin--$\frac{3}{2}$ 
pentaquarks should be sensible as well and are roughly expect between 
$1.6$ and $2.1{\rm GeV}$. 

\section*{Acknowledgments}
I am grateful to the organizers for providing this 
pleasant and worthwhile workshop. Furthermore I 
acknowledge interesting discussions with G. Holzwarth, 
R. L. Jaffe, J. Schechter, and H. Walliser.
This work is supported in parts by
DFG under contract We--1254/8--1.

\end{document}